\documentclass{pnastwo}

\usepackage{graphicx}
\usepackage[square,sort,comma,numbers]{natbib}

\newcommand{\ssr}{Space Sci. Rev.}
\newcommand{\nar}{New Astron. Rev.}
\newcommand{\apj}{Astrophys. J.}
\newcommand{\apjl}{Astrophys. J. Lett.}
\newcommand{\apjs}{Astrophys. J. Suppl.}
\newcommand{\aap}{Astron. Astrophys.}
\newcommand{\apss}{Astrophys. Space Sci.}
\newcommand{\mnras}{Mon. Not. R. Astron. Soc.}

\newcommand{\prc}{Phys. Rev. C}
\newcommand{\gca}{Geochim. Cosmochim. Acta}
\newcommand{\physrep}{Phys. Rep.}
\newcommand{\pasa}{Publ. Astron. Soc. Austr.}

\begin{document}

\title{Origin of the $p$-process radionuclides $^{92}$Nb and $^{146}$Sm
in the early Solar System and inferences on the birth of the Sun}


\author{Maria Lugaro\affil{1}{Konkoly Observatory, Research Centre for Astronomy and Earth Sciences,
Hungarian Academy of Sciences, H-1121 Budapest, 
Hungary}\affil{2}{Monash Centre for Astrophysics, Monash University, VIC3800, Australia},
Marco Pignatari\affil{1}{}\affil{3}{Nugrid collaboration},
Ulrich Ott\affil{4}{Faculty of Natural Science, University of West Hungary, H-9700 
Szombathely, Hungary}\affil{5}{Max-Planck Institute for Chemistry, D-55128 Mainz, 
Germany},
Kai Zuber\affil{6}{Institut f\"ur Kern- und Teilchenphysik, Technische Universit\"at
Dresden, D-01069 Dresden, Germany},
Claudia Travaglio\affil{7}{Osservatorio di Torino, I-10025, Pino Torinese, Italy}\affil{3}{},
Gy\"orgy Gy\"urky\affil{8}{Atomki Institute for Nuclear Research, 
Hungarian Academy of Sciences, H-4001, Debrecen, Hungary}, 
\and
Zsolt F\"ul\"op\affil{8}{}}

\contributor{Submitted to Proceedings of the National Academy of Sciences
of the United States of America}


\significancetext{
Radioactive nuclei with half lives of the order of million of years were present in 
the Solar System at 
its birth and can be used as clocks to measure the events that predated the birth of the Sun. Two of
these nuclei are heavy and rich in protons and can be produced only by particular chains of nuclear
reactions during some supernova explosions. We have used their abundances to derive that at least 10
million years passed between the last of these explosions and the birth of the Sun. This means that the
region were the Sun was born must have lived at least as long, which is possible only if it was very
massive.}


\maketitle

\begin{article}
\begin{abstract}

{The abundances of $^{92}$Nb and $^{146}$Sm in the early Solar System are determined from 
meteoritic analysis and their stellar production is attributed to the $p$ process. We investigate 
if their origin from thermonuclear supernovae deriving from the explosion of white dwarfs 
with mass above the Chandrasekhar limit is in agreement with the abundance of $^{53}$Mn, another 
radionuclide present in the early Solar System and produced in the same events. A consistent 
solution for $^{92}$Nb and $^{53}$Mn cannot be found within the current uncertainties and   
requires that the $^{92}$Nb/$^{92}$Mo ratio in the early Solar System is at least 50\% lower than 
the current nominal value, which is outside its present error bars. A different solution is to invoke 
another production site for $^{92}$Nb, which we find in the $\alpha$-rich freezeout during  
core-collapse supernovae from massive stars. Whichever scenario we consider, we find that a 
relatively long time interval of at least $\sim$ 10 Myr must have elapsed from when the 
star-forming region where the Sun was born was isolated from the interstellar medium and the birth 
of the Sun. This is in agreement with results obtained from radionuclides heavier than iron 
produced by neutron captures and lends further support to the idea that the Sun was born in a 
massive star-forming region together with many thousands of stellar siblings.} 
\end{abstract}




\dropcap{R}adionuclides with half lives of the order of 1 to 100 million years (Myr) were 
present in the early Solar System (ESS) and represent a clue to 
investigate the circumstances of the birth of the Sun \cite{dauphas11}. 
The abundances of some of these nuclei in
the ESS, as inferred from analysis of meteorites and
reported relative to the abundance of a stable nucleus, are 
given with error bars as low as a few percent. However, 
our ability to 
exploit radionuclides as chronometer for the events that led to the birth of the Sun
is undermined by the fact that our 
understanding of their production by nuclear reactions in stars is 
still relatively 
poor. This is because it relies on stellar models that include highly 
uncertain physics related to, e.g., mixing, the mechanisms of supernova explosions, 
and nuclear reaction rates. 
Here we focus on the origin of radionuclides heavier than Fe. Among 
them, $^{92}$Nb, $^{129}$I, $^{146}$Sm, and $^{182}$Hf 
have abundances in the ESS known with error bars lower than
20\%.
Lugaro {\it et al}. \cite{lugaro14} recently proposed a decoupled 
origin for $^{129}$I and $^{182}$Hf: The former was produced by $rapid$ 
neutron captures ($r$ process) in a supernova or neutron star merger 
roughly 100 Myr before the formation of the Sun, and the latter by 
$slow$ neutron captures ($s$ process) in an asymptotic giant branch star 
roughly 10--30 Myr before the formation of the Sun. From these timescales it was
concluded that the star forming region where the Sun was born may have lived for a few 
tens of million years, which is possible only if such a region was very massive and 
gave birth to many hundreds of stellar siblings \cite{murray11}.

Here, we turn to the investigation of the origin the other two radionuclides
heavier than Fe  
whose abundances are well known in the ESS: $^{92}$Nb and $^{146}$Sm. 
These nuclei are proton rich, relative to the stable nuclei of Nb and Sm, which 
means that they cannot be produced by neutron captures like the vast majority  
of the nuclei heavier than Fe. Instead, their nucleosynthetic 
origin has been traditionally ascribed to some flavor of the so-called 
$p$ process \cite{arnould03,rauscher13a}, for example, the 
disintegration of heavier nuclei ($\gamma$ process).
Unfortunately, the possible astrophysical sites of origin of 
$p$-process nuclei in the Universe 
are not well constrained. 

Core-collapse supernova explosions (CCSNe \cite{woosley02}) 
deriving from the final 
collapse of massive stars have been considered as a possible site for the 
$\gamma$ process, specifically, occurring in the O-Ne-rich zones of the 
CCSN ejecta \cite{woosley78}.
However, models never managed to reproduce the complete
$p$-process pattern observed in the bulk of the Solar System material 
\cite{prantzos90,rayet95,rauscher02}. For instance, CCSN models cannot reproduce 
the relatively large abundances of $^{92,94}$Mo and $^{96,98}$Ru. 
Taking into account nuclear uncertainties has not solved the problem \cite{rapp06,rauscher06}, 
except for a possible increase of the $^{12}$C+$^{12}$C fusion reaction rate 
\cite{bennett12,pignatari13b}. 
Another process in CCSNe that can produce the light $p$-process nuclei up to Pd-Ag, 
including $^{92}$Nb, is the 
combination of $\alpha$, proton, neutron captures, and their reverse reactions in $\alpha$-rich 
freezeout conditions \cite{woosley92}. Neutrino winds from the forming neutron star are
also a possible site for the production of the light $p$-process nuclei 
\cite{hoffman96,farouqi09,arcones11}, although, 
one of its possible components (the so-called $\nu$p-process \cite{froehlich06}) cannot produce 
$^{92}$Nb because it is shielded by $^{92}$Mo \cite{fisker09,rauscher13a}. The same occurs in the case 
of the {\it rp}-process in X-ray bursts \cite{dauphas03}. 
Neutrino-induced reactions in CCSNe (the $\nu$-process) can also produce some 
$^{92}$Nb \cite{hayakawa13}, but no other $p$-process nuclei. 

Thermonuclear supernovae (SNeIa) deriving from the explosion of a white dwarf that 
accreted material from a main sequence companion to above the Chandrasekhar limit 
have been proposed as a site of the 
$\gamma$ process \cite{kusakabe11,travaglio11,travaglio15}. 
In these models, heavy seed nuclei are assumed to be produced by the $s$ process
during the accretion phase and, given a certain initial $s$-process distribution, 
it is possible to reproduce the high abundances of 
$^{92,94}$Mo and $^{96,98}$Ru in the Solar System. 
However, there are still many uncertainties related to 
the origin and the features of the 
neutron-capture processes activated during the white dwarf accretion
leading to SNIa. 
Recently, Travaglio {\it et al}. \cite{travaglio14} (hereafter TR14) analyzed in detail the production of 
$^{92}$Nb and $^{146}$Sm in SNeIa using multidimensional models 
and concluded that such an origin is plausible for 
both radionuclides in the ESS. Here, we consider these results and 
combine them with new predictions of the production of $p$-process nuclei in 
$\alpha$-rich
freezeout conditions in CCSNe to investigate the 
origin of $^{92}$Nb and $^{146}$Sm and to use it to further constrain the circumstances of the 
birth of the Sun.

\section{Abundance ratios and related timescales}
\label{sec:ratios}

In simple analytic terms the 
abundance ratio of a short-lived (T$_{1/2} <$ 100 Myr) radioactive to a stable
isotope in the material that ended up in the Solar System, 
just after a given nucleosynthetic event and 
assuming that both nuclides are only produced by this
type of event, can be derived using the formula \cite{wasserburg06,lugaro14}:
\begin{equation} 
\label{eq:eq1} 
\frac{\mathrm N_{radio}}{\mathrm N_{stable}} = {\mathrm K} \times \frac{\mathrm p_{radio}}{\mathrm 
p_{stable}} \times \frac{\delta}{\mathrm T} \times \left(1 + 
\frac{e^{-\delta/\tau}}{1 - e^{- \delta/\tau}}\right), 
\end{equation} 
where ${\mathrm p_{radio}/\mathrm p_{stable}}$ is the production ratio of each 
single event, $\tau$ is the mean life (=T$_{1/2}$/ln 2) of the radionuclide, 
${\mathrm T}$ the timescale of the evolution of the Galaxy up to the formation of the Sun 
($\sim10^{10}$ yr), and $\delta$ the 
recurrence time between each event. The value of $\delta$ 
is essentially a free parameter that may vary between 10 
and 100 Myr \cite{meyer00}. The number K is also a free parameter that 
accounts for the effect of infall of low-metallicity 
gas, which dilutes the abundances, and the fact that a fraction of the abundances produced, 
particularly for stable isotopes, is locked inside stars \cite{clayton85}. 
The value of K changes depending on whether the isotopes involved are of primary or secondary origin, i.e., 
whether they are produced from the H and He abundances in the star or depend on the initial 
presence of CNO elements, respectively. These effects are complex to evaluate analytically, 
but the general result is that K $>$ 1 \cite{huss09}. Lugaro {\it et al}.
\cite{lugaro14} did not consider the number K in their evaluation of ratios from Eq.~\ref{eq:eq1}, 
which means that effectively they used K=1 and their reported timescales represent conservative 
lower limits.

TR14 did not use Eq.~\ref{eq:eq1}, but included the yields of $^{92}$Nb, $^{97}$Tc, $^{98}$Tc, and 
$^{146}$Sm (and their reference isotopes $^{92}$Mo, $^{98}$Ru, and $^{144}$Sm) from their 
SNIa models into full, self-consistent Galactic chemical evolution (GCE) simulations to evaluate 
the abundance ratios $^{92}$Nb/$^{92}$Mo, $^{97}$Tc/$^{98}$Ru, $^{98}$Tc/$^{98}$Ru, and 
$^{146}$Sm/$^{144}$Sm in the interstellar medium (ISM) at the time of the birth of the Sun, assuming 
that the production of $p$ nuclei only occurs in SNIa. These detailed models reproduce
the abundances of the stable reference isotopes considered here \cite{travaglio15}.
With Eq.~\ref{eq:eq1} we can recover results close to those of the detailed GCE models for the four ratios 
considered here using as
${\mathrm p_{radio}/\mathrm p_{stable}}$ the average of the values given in
Table~1 of TR14\footnote{We average only the values given for metallicities in the range 
$Z$ = 0.01 -- 0.02, since we are focusing on
events that occurred close in time to the formation of the Sun. Variations in this range of
$Z$ are within 25\%. When considering also
$Z$ down to 0.003 they are within a factor of 2. The only exception is $^{98}$Tc/$^{98}$Ru, which
varies by 50\% in the range $Z$ = 0.01 -- 0.02 and by a factor of 6 when also
considering $Z$ = 0.003.}, T = 9200 Myr from TR14, $\delta$ = 8 Myr, and K$ = 2$ (Table~\ref{tab:tab}).
This means that roughly (1/K)(T/$\delta) \simeq$ 600 SNIa $p$-process events contributed to 
the Solar System abundances of the stable isotopes considered here. For the unstable isotopes, 
it depends on their mean life: because $^{97}$Tc and $^{98}$Tc have relatively short mean 
lives, the second term of the sum in Eq.~\ref{eq:eq1} representing 
the memory of all the events prior the last event counts for 26\% of the total, hence, most
of their ESS abundances come from the last event. 
On the other hand, for $^{92}$Nb and $^{146}$Sm, due to their long half 
lives, the second term of the sum is the most important. For example, in the case of 
$^{92}$Nb it accounts for 85\% of the total amount of $^{92}$Nb.
In these cases the ratios from Eq.~\ref{eq:eq1} 
are very close to the 
corresponding ISM steady-state ratio, i.e., the production ratio multiplied by K$\tau$/T. 

Although we can recover the values of the ratios produced by the full GCE models using Eq.~\ref{eq:eq1}, 
we need to keep in mind some distinctions. The ratios derived by TR14 using 
the full GCE model represent values at an absolute time 9200 Myr from the birth of the Galaxy, 
when the ISM reaches solar metallicity.
From these values, 
we can evaluate an isolation timescale (T$_{\rm isolation}$): the interval between 
the time when the material that ended up in the Solar System became isolated from the 
ISM and the time of the formation of the Solar System. 
T$_{\rm isolation}$ is derived such that the ratio between the ESS ratio and the ISM ratio for a 
given radioactive and stable pair is simply given 
by $e^{-{\rm T}_{\rm isolation}/\tau}$. 
In reality, however, 
some mixing could have occurred. An ISM mixing timescale (T$_{\rm mixing}$) 
between different phases 
of the ISM should be of the order of 10 - 100 Myr. The effect of such process  
was analytically investigated by \cite{clayton83}, from which 
\cite{rauscher13a} derived that the ratio between the ESS ratio and the ISM ratio for a given 
radioactive and stable pair is given by 
$1 + 1.5 x + 0.4 x^2$, where $x =$ T$_{\rm mixing}/\tau$. 
In this picture, the requirement is that the composition of the star-forming region where 
the Sun was born must have been affected by mixing with the ISM, hence, T$_{\rm mixing}$ represents 
a lower limit for T$_{\rm isolation}$. 

Values derived using Eq.~\ref{eq:eq1}, instead, represent ratios in the matter that 
built up the Solar System just after the last, final addition from a nucleosynthetic event. From 
them, we can evaluate a last-event timescale (T$_{\rm last}$): the interval between the 
time of the last event and the time of the formation of the Solar System. T$_{\rm last}$ 
represents an upper limit of T$_{\rm isolation}$ and 
is derived such that the ratio between the ESS ratio and the ISM ratio for a 
given radioactive and stable pair is simply given by $e^{-{\rm T}_{\rm last}/\tau}$ (as for 
T$_{\rm isolation}$). The more 
$\delta/\tau$ is lower than unity, the closer the ratio derived from Eq.~\ref{eq:eq1} is 
to the ISM ratio, and the closer T$_{\rm last}$ to T$_{\rm isolation}$.
The main drawback of this approach is that the K and 
the $\delta$ values in Eq.~\ref{eq:eq1} are uncertain. The advantages are that 
there are not further complications with regards to mixing in the ISM 
and that this description is relatively free from uncertainties related to the setting of the time of the 
birth of the Sun. In the model of TR14, this is defined as the time when the Galactic metallicity 
reaches the solar value. However, stellar ages and metallicities in the Galaxy are not strictly 
correlated 
\cite{casagrande11}. Chemodynamical GCE models also including the effect of stellar migration 
\cite{sanchez09,kobayashi11} are required to obtain the observed spread, but have not been 
applied to the evolution of radioactive nuclei yet. We expect these 
models to produce a range of abundances of the radioactive nuclei at any given 
metallicity, but we cannot predict based on first principles if these variations will be significant.  
Finally, it should be noted that, while T$_{\rm isolation}$ and T$_{\rm mixing}$ are required  
to be the same for all the different types of nucleosynthetic events 
that contributed to the Solar System matter, T$_{\rm last}$
is in principle different for each type of event. 

\section{The SNIa source}

We evaluate T$_{\rm isolation}$ and T$_{\rm mixing}$ assuming that the  
$p$-process radionuclides were produced by SNeIa and comparing 
the abundance ratios obtained by TR14 to those observed in the ESS (Table~\ref{tab:tab}).
TR14 carefully analysed the nuclear 
uncertainties that affect the production of $^{92}$Nb and $^{146}$Sm in SNeIa. 
In the production of $^{92}$Nb, ($\gamma$,n) reactions play the dominant role with some 
contribution from proton induced reactions. The nuclear uncertainties are related to 
the choice of the $\gamma$-ray strength function and, to a lesser extent, the proton-nucleus 
optical potential. Since the rates of some of the most important reactions are constrained  
experimentally, the nuclear uncertainties on the $^{92}$Nb production are moderate, resulting 
in possible variations in the ISM ratio of less than a factor of two. 
The $^{146}$Sm/$^{144}$Sm ratio, on the other hand, is determined by 
($\gamma$,n)/($\gamma$,$\alpha$) branchings, mainly on $^{148}$Gd. 
The uncertainty range reported by TR14 is 
based on three choices of the $^{148}$Gd($\gamma$,$\alpha$)$^{144}$Sm rate. 
With respect to two previous estimates \cite{somorjai98,rauscher00}, the new rate of  
\cite{rauscher13b} results in a $^{146}$Sm/$^{144}$Sm ratio higher by a factor of two 
in SNIa, but lower by at least a factor of two in CCSNe \cite{rauscher13b}.
Realistically, the $^{146}$Sm/$^{144}$Sm ratio 
may have an even higher nuclear uncertainty, possibly up to one order of magnitude,
owing to the lack of experimental data at the relevant low energies for the 
$^{148}$Gd($\gamma$,$\alpha$)$^{144}$Sm rate 
itself and for the low energy $\alpha$-nucleus optical potential required 
for the extrapolation \cite{gyurky14}.
For $^{92}$Nb/$^{92}$Mo, if we compare
the maximum value allowed within nuclear uncertainties of 3.12 $\times 10^{-5}$ 
to the lower limit of the ESS value, we derive a maximum T$_{\rm isolation}$ of 5.4 Myr and a maximum 
T$_{\rm mixing}$ of 3.7 Myr. It is also possible to recover a solution for $^{146}$Sm/$^{144}$Sm 
consistent with these values, given the large uncertainties. Currently, the value of the 
half life 
of $^{146}$Sm is debated between the two values of 68 and 103 Myr  
\cite{kinoshita12,marks14}. Here, we have used the lower value, if we used the higher value
we would obtain longer timescales than those reported in Table~\ref{tab:tab}.

We extend the study of TR14 to investigate if the origin of $^{92}$Nb and $^{146}$Sm 
is compatible with that of the radionuclide $^{53}$Mn. This nucleus provides us a  
strong constraint because near Chandrasekhar-mass SNeIa are also the major producers of Mn 
in the Solar System \cite{seitenzahl13} and, together with the stable $^{55}$Mn, they produce 
$^{53}$Mn, whose solar abundance is well determined in the ESS 
(Table~\ref{tab:tab}). We calculate the ISM $^{53}$Mn/$^{55}$Mn
ratio from Eq.~\ref{eq:eq1} using the same values for
$\delta$, K, and T derived above and  
the production ratio $^{53}$Mn/$^{55}$Mn=0.108 predicted by the same SNIa model 
used to derive the $p$-process abundances. This is derived from the $^{53}$Cr abundance 
given by \cite{travaglio11}, but, we note that different models produce very similar 
$^{53}$Mn/$^{55}$Mn ratios \cite{travaglio04}. We calculate an ISM 
$^{53}$Mn/$^{55}$Mn ratio of 2.41 $\times 
10^{-4}$, which, after an isolation time of 
5.4 Myr, results in a ratio of 5.82 $\times 10^{-5}$, one order of magnitude higher 
than the ESS value. An isolation time of at least 19 Myr is instead required to 
match the $^{53}$Mn/$^{55}$Mn constraint.
If we make the conservative assumption that K=1, 
instead of 2, to calculate the ISM $^{53}$Mn/$^{55}$Mn 
ratio from Eq.~\ref{eq:eq1} we obtain T$_{\rm isolation} \simeq$ 15 Myr.
This value is strongly incompatible with the upper limit of 5.4 Myr 
required to match the $^{92}$Nb/$^{92}$Mo, in other words, SNIa 
nucleosynthesis results in too high production of $^{53}$Mn relative to $^{92}$Nb, and to 
their ESS abundances.

A three times lower $^{53}$Mn/$^{55}$Mn ratio in SNIa would result from a tenfold increase of the 
$^{32}$S($\beta^+$)$^{32}$P decay \cite{parikh13}, in which case T$_{\rm isolation}$ would decrease 
to $\simeq$ 9 Myr. This possibility needs to be further investigated. On the other hand, a possibly 
longer half life for $^{53}$Mn (see discussion in \cite{dressler12}) would further increase the 
difference. To reconcile the ESS $^{92}$Nb abundance with an isolation 
time of 15 Myr, the half life of $^{92}$Nb should be almost a factor of three higher,
which seems unrealistic. The current half life is the 
weighted average of two experiments that produced similar results in spite of being based on 
different normalizations: the first \cite{makino77} is normalized to the half life of $^{94}$Nb, 
which is not 
well known, and the second \cite{nethaway78} to an assumed value of the $^{93}$Nb(n,2n)$^{92}$Nb cross 
section. Another option is that the $^{92}$Nb/$^{92}$Mo ratio in the ESS is lower than 
2.2 $\times 10^{-5}$, i.e., 21\% lower than the current lower limit.
It is harder to reconcile the different T$_{\rm mixing}$: when using the lowest (K=1) ISM 
$^{53}$Mn/$^{55}$Mn ratio and the upper limit for the ESS ratio, we obtain
a lower limit for T$_{\rm mixing}$ of 24 Myr. To reconcile 
the $^{92}$Nb/$^{92}$Mo
ratio to this timescale an ESS $^{92}$Nb/$^{92}$Mo=1.6$\times 10^{-5}$ is required, 
43\% lower than the current lower limit. A different solution is to look into another production site for
$^{92}$Nb.

\section{The CCSN source}

\begin{figure*}
\center{\includegraphics[scale=0.7,angle=270]{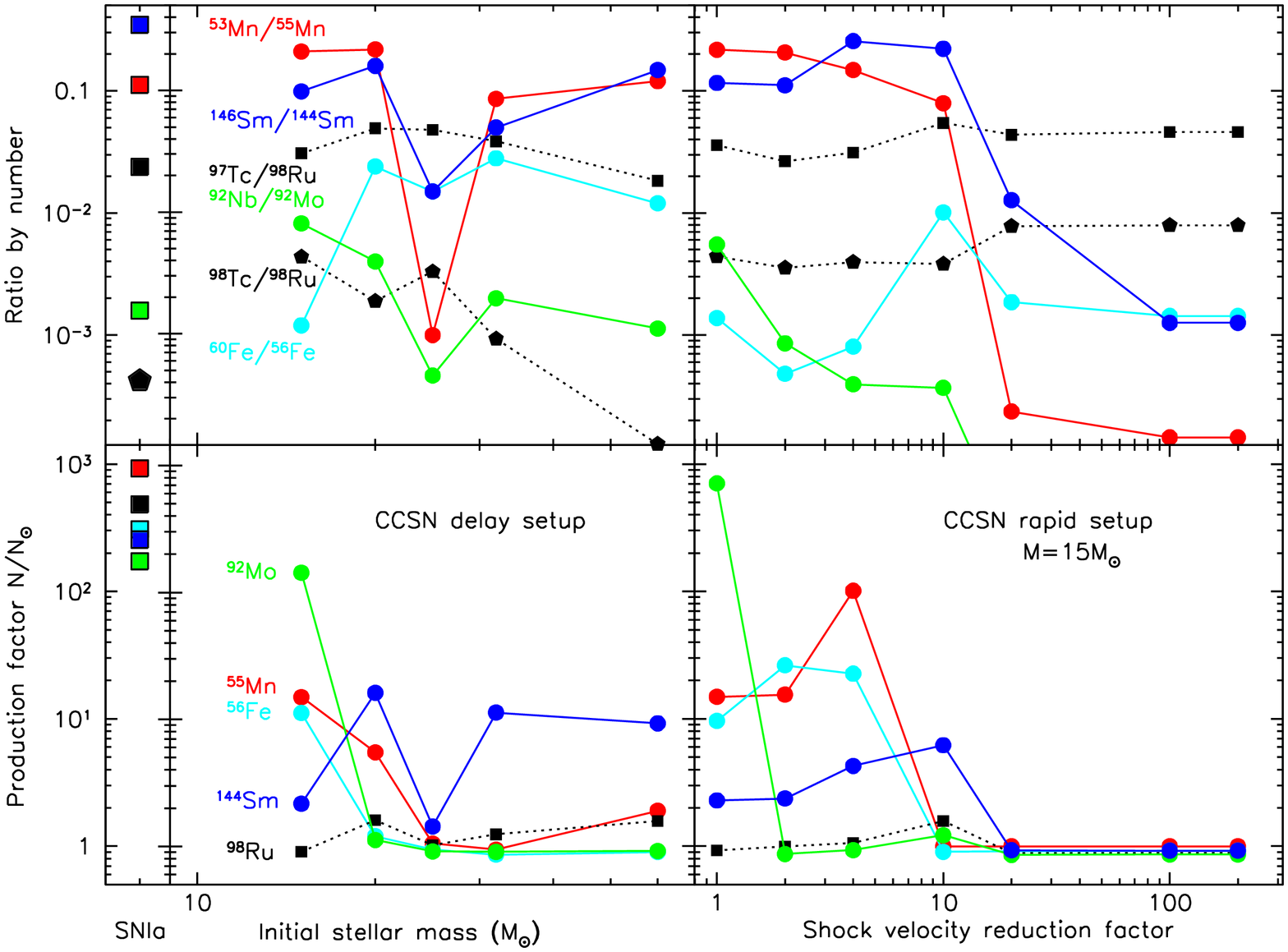}}
\caption{Results for the radioactive and stable nuclei of interest from the CCSN models
as compared to those from SNIa (TR14). 
The CCSN models with the delay setup have initial masses 15, 20, 
25, 32 and 60 M$_{\odot}$ and metallicity
0.02. The CCSN models with the rapid setup are for a star of initial mass 15 M$_{\odot}$
and different reduction factors for the shock velocity. 
The rate by \cite{rauscher00} was used for the $^{148}$Gd($\gamma$,$\alpha$)$^{144}$Sm reaction. 
As mentioned in the text, using the new rate by \cite{rauscher13b} 
would result in $^{146}$Sm/$^{144}$Sm ratios higher in SNIa and 
lower in CCSNe.}  
\label{fig:ratios}
\end{figure*}

As mentioned above, the $\gamma$-process in CCSN models do not efficiently produce 
$p$-process isotopes in the Mo-Ru region, but, other CCSN nucleosynthesis components 
may contribute to these isotopes. Pignatari {\it et al}. 2013 \cite{pignatari13a} computed 
CCSN models with initial mass of 15 M$_{\odot}$ that carry in the ejecta an $\alpha$-rich freezeout 
component. There, production of the $p$-process nuclei up to $^{92}$Mo occurs due to a combination of 
$\alpha$, proton, and neutron captures and their inverse reactions during the CCSN explosion in the 
deepest part of the ejecta, where $^4$He is the most abundant isotope. 
The stellar evolution before the CCSN explosion was calculated with the code GENEC 
\cite{eggenberger08} and the explosion simulations were based on the prescriptions for shock 
velocity and fallback by \cite{fryer12}. The standard initial shock velocity used 
beyond fallback is $2 \times 10^9$ cm s$^{-1}$ for all the masses, and
we also experimented by reducing it down to 200 times lower.
Results are presented 
for CCSN models computed with two setups for the convection-enhanced neutrino-driven explosion: 
fast-convection (the rapid setup) and delayed-convection 
(the delay setup) explosions \cite{fryer12}.
In this framework, nuclear uncertainties are insignificant compared to 
model uncertainties. 
A post-processing code was used to calculate the nucleosynthesis before and during the explosion 
\cite{pignatari12}. The results are summarized and 
compared to the SNIa models in Figure~\ref{fig:ratios}.

In comparison to the SNIa models, the CCSN models produce less 
$^{53,55}$Mn, $^{56}$Fe, and $^{144,146}$Sm, and almost no $^{98}$Ru and $^{97,98}$Tc.
On the other hand,
no $^{60}$Fe is produced in SNeIa, and 
the cosmic origin of this isotope
has to be ascribed to other types of supernovae, including CCSNe.
Significant production of $^{92}$Nb and $^{92}$Mo
occurs in CCSN models of 15 M$_{\odot}$, with production factors for
$^{92}$Mo comparable to those of the SNIa 
models, and $^{92}$Nb/$^{92}$Mo ratios up to five times higher.
CCSN models with initial mass larger than 15 M$_{\odot}$ do not eject material exposed to the 
$\alpha$-rich freezeout due to the more extended fallback. 
The $\alpha$-rich freezeout efficiency also strongly depends on the shock velocity.  
When its 
standard value is reduced, even only by a factor of two, the amounts of $^{92}$Mo and $^{92}$Nb
ejected become negligible. This is because for lower shock velocities the bulk of 
$\alpha$-rich freezeout nucleosynthesis is shifted toward lighter elements closer to the Fe 
group. Only some $\gamma$-process component is produced in these cases, which is poor in $^{92}$Mo. 
In summary,  
the $\alpha$-rich freezeout conditions suitable to produce $^{92}$Mo and $^{92}$Nb
are more likely hosted in CCSNe
with initial mass of 15 M$_{\odot}$ or lower and  
shock velocities at least as large as those 
provided by \cite{fryer12}. 

We explored the possibility that the $^{92}$Nb in the ESS came from the $\alpha$ freezeout 
production experienced by these 15 M$_{\odot}$ CCSN models. 
Since the premise of Eq.~\ref{eq:eq1} is that both 
the stable and unstable isotopes present in the ESS were produced only by this type of event, 
under such assumption, we can apply Eq.~\ref{eq:eq1} 
only to calculate the $^{92}$Nb/$^{92}$Mo ratios to derive 
the time of the last such $\alpha$ freezeout event, since the other isotopes are not 
produced in any significant abundances\footnote{Even if all the Mn came from these events we 
would obtain timescales consistent with those derived from $^{92}$Nb.}. 
The calculations are hampered by the error bars on the ESS ratios and 
the large uncertainties related to the choice of the 
free parameters entering in Eq.~\ref{eq:eq1}. However, to infer further 
information on the birth of the Sun we are particularly interested in determining the lower 
limit of T$_{\rm last}$ because this provides us the shortest possible timescale, within all the 
uncertainties above, for the life of the 
star forming region where the Sun was born. 
Using K=1 in Eq.~\ref{eq:eq1} and the upper limit of the ESS $^{92}$Nb/$^{92}$Mo ratio 
we inferred a lower limit for
T$_{\rm last}$ of 10 Myr when using the delay setup CCSN model (Table~\ref{tab:tab2}).
However, we stress that K=1 represents a very conservative lower limit 
because the production factor $^{92}$Mo/$^{92}$Mo$_{\odot}$ of the CCSN delay model is 
comparable to that of the SNIa model (140 versus 172). This means that we would need a 
similar number of events (which scales as 1/K$\delta$) to 
reproduce the $^{92}$Mo solar abundance - although the exact value would depend 
on the metallicity dependence of $^{92}$Mo/$^{92}$Mo$_{\odot}$ in the CCSN models, 
which we did not explore. 
The lower limit of T$_{\rm last}$ from the rapid setup CCSN model 
is instead a negative value. However, 
the production factor $^{92}$Mo/$^{92}$Mo$_{\odot}$ of the rapid model is 
five times higher than that of the delay model (704 versus 140).
This means that we would need five
times less events to reach the same value of $^{92}$Mo in the Solar System, hence, for consistency, 
a five times higher value of 1/K$\delta$ is more appropriate 
in this case. If we use, e.g., $\delta$= 50 Myr we obtain 7 Myr as the convervative 
lower limit for T$_{\rm last}$. 

\section{Discussion}

Our models show that $^{146}$Sm in the ESS was produced by SNIa events, together with 
$^{97,98}$Tc. The large nuclear uncertainties on the former and the availability of 
upper limits only for the ESS ratio for the latter two do not allow us to derive precise timescales.
On the other hand, to reconcile the abundances of $^{53}$Mn and $^{92}$Nb in the ESS 
as produced primarily by SNIa events, the ESS $^{92}$Nb/$^{92}$Mo ratio needs to be 
at least 50\% lower than the 
current nominal value, which is outside its present error bars. 
In this case
T$_{\rm isolation}$ $\simeq$ 15 Myr, which is consistent with the T$_{\rm last}$ derived for the  
last $r$ ($\sim$ 100 Myr) and $s$ ($\sim$ 20 Myr) process events \cite{lugaro14} since 
T$_{\rm last}$ represents an upper limit for T$_{\rm isolation}$. 
Furthermore, if all the $^{53}$Mn in the ESS was produced by the same 
SNIa events that produced $^{92}$Nb, no CCSNe exploding within the star-forming 
region where the Sun was born should have contributed to the abundances of radioactive nuclei in 
the ESS, since they would have contributed extra $^{53}$Mn. If correct, this will have important 
consequences for the origin of $^{26}$Al and $^{60}$Fe in the ESS. 
The $^{60}$Fe/$^{56}$Fe is likely  
the result of the decay, after a certain isolation time, of the $^{60}$Fe abundance in the ISM. 
This is consistent with the direct observational constraint from $\gamma$-ray astronomy 
of $^{60}$Fe/$^{56}$Fe $\simeq 2.8 \times 10^{-7}$ in the ISM, and with timescales of the 
order of 15 Myr, while $^{26}$Al may have originated from the winds of a massive star \cite{tang12}. 

Another possibility is that the $^{92}$Nb was instead predominantly produced by CCSNe that experienced 
the $\alpha$-rich freezeout, corresponding in our models to low initial stellar masses experiencing a 
core-collapse with high shock velocities. In the extreme scenario where all the $^{92}$Nb and 
$^{92}$Mo in the ESS came from such events we have derived lower limits for the timing of the last of 
them. These are also still consistent with a relatively long T$_{\rm isolation}$ of at least 7 
Myr, since T$_{\rm isolation}$ $\sim$ T$_{\rm
last}$ when $\delta << \tau$, which is verified for $^{92}$Nb when $\delta$=10 Myr.

Clearly, $^{92}$Nb could have been produced by both SNIa and CCSN events. Full GCE models are 
required, including both SNIa, the $\alpha$-rich freezeout component of CCSN considered here, and the 
possible production in the neutrino winds.
Because the $\alpha$-rich freezeout CCSNe events of initial mass 15 M$_{\odot}$ produce 
higher amounts of $^{92}$Nb than SNIa, we 
expect that adding them to the balance will increase the ISM $^{92}$Nb/$^{92}$Mo 
ratio at the time of 
the formation of the Solar System, leading to longer T$_{\rm isolation}$, and an easier 
match with the 
$^{53}$Mn/$^{55}$Mn constraint. In conclusion, a 
higher-precision determination of the ESS $^{92}$Nb/$^{92}$Mo ratio 
is urgently required. This will allow us to determine if SNIa are the major source of 
light $p$-process 
nuclei in the Galaxy or if CCSNe also play a role; and which fraction of the $^{53}$Mn 
in the ESS originated from SNIa.



\begin{acknowledgments} ML is a Momentum (``Lend\"ulet-2014'' Programme) project leader of the Hungarian 
Academy of Sciences. MP acknowledges support to NuGrid from NSF grants PHY 09-22648 (Joint Institute for 
Nuclear Astrophysics, JINA), NSF grant PHY-1430152 (JINA Center for the Evolution of the Elements) and EU 
MIRGCT-2006-046520. MP acknowledges the support from the ``Lend\"ulet-2014'' Programme of the Hungarian 
Academy of Sciences (Hungary) and from SNF (Switzerland). \end{acknowledgments}

\small

\end{article}

\begin{table*}[h]
\caption{Decay rates, ESS ratios, SNIa production and ISM ratios, and 
timescales for the isotopes of interest}\label{tab:tab}
\begin{tabular}{l l l l l l l}
 & $^{\mathsf 53}$Mn & $^{\mathsf 60}$Fe & $^{\mathsf 92}$Nb & $^{\mathsf 97}$Tc & $^{\mathsf 98}$Tc 
& $^{\mathsf 146}$Sm \\ 
\hline
T$_{\mathsf 1/2}$(Myr) & 3.7 & 2.6 & 35 & 4.2 & 4.2 & 68 \\  
$\tau$(Myr) & 5.3 & 3.8 & 50 & 6.1 & 6.1 & 98 \\
reference isotope & $^{\mathsf 55}$Mn & $^{\mathsf 56}$Fe & $^{\mathsf 92}$Mo & $^{\mathsf 98}$Ru & 
$^{\mathsf 98}$Ru & $^{\mathsf 144}$Sm \\  
ESS ratio & (6.28$\pm$0.66)$\times$10$^{-6}$ & (5.39$\pm$3.27)$\times$10$^{-9}$
& (3.4$\pm$0.6)$\times$10$^{-5}$ & $<$ 4$\times$10$^{-5}$ & $<$ 6$\times$10$^{-5}$ & (9.4$\pm$0.5)$\times$10$^{-3}$ \\       
reference & \cite{trinquier08}\tablenote{The
most recent value of 6.8$\times$10$^{-6}$ given by \cite{gopel15} is included in the
given error bars.} & \cite{tang15} & \cite{harper96,lodders09} & \cite{dauphas02,lodders09} & \cite{becker03,lodders09} & 
\cite{boyet10,travaglio14} \\
production ratio\tablenote{Average of the ratios from Table 1 of TR14 from metallicities 
from 0.01 to 0.02, except for $^{\mathsf 53}$Mn/$^{\mathsf 55}$Mn from \cite{travaglio11} 
and $^{\mathsf 60}$Fe/$^{\mathsf 56}$Fe from \cite{travaglio04}.}
 & 0.108 & $<$ 10$^{\mathsf -9}$ & 1.58$\times$10$^{\mathsf -3}$ & 2.39$\times$ 10$^{\mathsf -2}$ 
    & 4.22$\times$10$^{\mathsf -4}$ & 0.347 \\  
ISM from GCE\tablenote{From Tables 2, 3, and 4 of TR14} & & 
& 1.72$^{+1.40}_{-0.06}\times$10$^{-5}$ & 4.08$\times$10$^{-5}$ 
& 6.47$\times$10$^{-7}$ & 7.0$^{+9.7}\times$10$^{-3}$ \\  
ISM from Eq.~\ref{eq:eq1}\tablenote{Using K=2, ${\mathsf \delta=8}$ Myr, and T=9200 Myr.} 
& 2.41$\times$10$^{-4}$ & $<$ 10$^{-12}$ & 1.86$\times$10$^{-5}$ 
    & 5.68$\times$10$^{-5}$ & 1.00$\times$10$^{-6}$ & 7.70$\times$10$^{-3}$ \\
T$_{\rm isolation}$\tablenote{Reported ranges reflect the error bars on the ESS and ISM ratios.}
 & 19 -- 20 & - & $\leq$ 5.4 & $>$ 0.12 & - & $\leq$ 62 \\
T$_{\rm mixing}$ & 40 -- 45 & - & $\leq$ 3.7 & $>$ 0.08 & - & $\leq$ 50 \\   
\hline
\end{tabular}
\end{table*}

\begin{table*}[h]
\caption{$^{\mathsf 92}$Nb/$^{\mathsf 92}$Mo from the CCSN models of 15 M$_{\odot}$, 
ISM ratios, and inferred lower limits for T$_{\rm last}$} 
\label{tab:tab2}
\begin{tabular}{l l l l l l l}
 & & delay setup & & & rapid setup & \\
\hline
production ratios & & 8.20$\times$10$^{-3}$ & & & 5.30$\times$10$^{-3}$ & \\
\hline
 & $\delta$ = 10 Myr & $\delta$ = 50 Myr & $\delta$ = 100 Myr & $\delta$ = 10 Myr & $\delta$ = 50 Myr & $\delta$ = 100 Myr \\
ISM ratio\tablenote{From Eq.~\ref{eq:eq1} using T=9200 Myr and K=1 to obtain a conservative 
lower limit.}
 & 4.92$\times$10$^{-5}$ & 7.06$\times$10$^{-5}$ & 1.03$\times$10$^{-4}$ &
3.18$\times$10$^{-5}$ & 4.56$\times$10$^{-5}$ & 6.66$\times$10$^{-5}$ \\
lower limit of T$_{\rm last}$\tablenote{Using the ESS upper limit of 4.0$\times$10$^{-5}$.} 
& 10 & 29 & 48 & - & 7 & 25 \\
\hline
\end{tabular}
\end{table*}

\end{document}